\documentclass[preprint,review]{elsarticle}

\usepackage{graphicx}
\usepackage{amssymb}
\usepackage{natbib}
\usepackage{lineno}
\usepackage{multirow}
\usepackage{hhline}
\usepackage{subfigure}

\usepackage{fancyhdr}
\fancypagestyle{pprintTitle}{%
  \lhead{}\lfoot{
    DOI: https://doi.org/10.1016/j.nima.2016.08.035 \\
 \ \\
    \copyright \ 2016. This manuscript version is made available under the CC-BY-NC-ND 4.0 license http://creativecommons.org/licenses/by-nc-nd/4.0/}\cfoot{}\rfoot{}
  
}
\journal{NIMA}

\begin{document}
\newcommand{\Angst}{$\mathring{\mathrm{A}}$}

\begin{frontmatter}

\title{Population-based metaheuristic optimization in neutron optics and shielding design}

\author[a,b]{D. D. DiJulio\corref{cor1}}
\ead{Douglas.DiJulio@esss.se}

\author[a,c]{H. Bj\"{o}rgvinsd\'{o}ttir}

\author[a]{C. Zendler}

\author[a,c]{P. M. Bentley}

\cortext[cor1]{Corresponding Author}

\address[a]{European Spallation Source ERIC, P.O. Box 176, SE-221 00 Lund, Sweden}
\address[b]{Division of Nuclear Physics, Lund University, SE-221 00 Lund, Sweden} 
\address[c]{Department of Physics and Astronomy, Uppsala University, Sweden}

\begin{abstract}
Population-based metaheutristic algorithms are powerful tools in the design of neutron scattering instruments and the use of these types of algorithms for this purpose is becoming more and more commonplace. 
Today there exists a wide range of algorithms to choose from when designing an instrument and it is not always initially clear which may provide the best performance. 
Furthermore, due to the nature of these types of algorithms, the final solution found for a specific design scenario cannot always be guaranteed to be the global optimum. Therefore, to explore the potential benefits and differences between the 
varieties of these algorithms available, when applied to such design scenarios, we have carried out a detailed study of some commonly used algorithms. For this purpose, we have developed a new general optimization software package 
which combines a number of common metaheuristic algorithms within a single user interface and is designed specifically with neutronic calculations in mind. The algorithms included in the software are implementations
of Particle-Swarm Optimization (PSO), Differential Evolution (DE), Artificial Bee Colony (ABC), and a Genetic Algorithm (GA). The software has been used to optimize the design of several problems
in neutron optics and shielding, coupled with Monte-Carlo simulations, in order to evaluate the performance of the various algorithms. Generally, the performance of the algorithms depended on the specific
scenarios, however it was found that DE provided the best average solutions in all scenarios investigated in this work. 
\end{abstract}

%
%


\hyphenation{}

\end{frontmatter}


\section{Introduction}

The use of population-based metaheuristic algorithms in the design of neutron scattering instruments 
is becoming more and more commonplace. These types of algorithms are well suited for the task, 
due in part to the large number of parameters involved in a typical design scenario and the resulting 
noisy parameter spaces. These two points make it often difficult to apply traditional optimization
algorithms for the design of such systems \cite{Lieutenant2005}.
Furthermore, the optimization of such an instrument is frequently a tedious and complex procedure. Automated
algorithms, which can efficiently search the parameter space with little user input, provide a great advantage
in this process. \\
\indent In a number of previous studies, population-based algorithms have been successfully applied to neutron optics and shielding
design. For example, a Genetic Algorithm (GA) \cite{GA} was used for the design of specific instruments at
the Hahn-Meitner Institute, Berlin and the Institut Laue-Langevin, France \cite{Bentley2006,Holzel2006}.
A GA was also used to optimize the composition of shielding material for mixed neutron and photon fields \cite{Hu2008}.
In some additional studies, Particle-Swarm Optimization (PSO) \cite{PSO} was applied to the design of an entire neutron guide hall \cite{Bentley2011}
and to multi-channel focusing guides \cite{Bentley2009}, where it showed improved performance over GA, and it was also used in the design
of elliptic focusing guides \cite{Damian}. Artificial Bee Colony (ABC) \cite{ABC} and Differential Evolution (DE) \cite{DE} were also used in the design of
multi-channel focusing guides for extreme sample environments \cite{DiJulio2014}. These algorithms demonstrated
improved performance compared to PSO for certain design scenarios. The benefits of population-based algorithms
were also noted in Ref. \cite{Farhi2014}, where a number of algorithms  were tested on a set of 
multi-dimensional objective functions where the global minimina were known. \\
\indent While metaheuristic algorithms have exhibited exceptional performance for the design of neutron
instruments, it can be expected that there may be noticeable differences between the results of individual algorithms when applied to the same design scenario. 
In practice, often only one type of algorithm is used for the design of a particular system. As the usage of metaheuristic algorithms cannot guarantee
that the global optimum will be found, it can be of great benefit to repeat the optimization with a number of different algorithms and starting conditions.
To explore this further, we have developed software containing a suite of algorithms combined under a single user interface and with neutron scattering applications in mind. The software is general
enough that it can be easily coupled to an external simulation package. In this work, we have coupled
the software to VITESS \cite{VITESS} for neutron optics calculations and Geant4 \cite{GEANT4,Geant4Ref1} for neutron shielding calculations
and applied it to several different design scenarios in order to evaluate the performance of each algorithm. In the following, we first provide a description
of the software developed, followed by the results of the applications of the software to the above mentioned type of calculations. Lastly, we discuss and present
the conclusions of our work.\\

\section{Description of the software}
The newly developed software contains implementations of PSO, GA, ABC, and DE. The ABC and DE packages are based on the freely downloadable codes from Ref. \cite{ABCsoftware,DEsoftware}.
The algorithms are described in detail below and an outline of the of the structure of the software is indicated in Figure 1. The parameters to be optimized, here refereed to as $\omega$, are specified in a parameter file which is read by the software. These parameters can be either discreet or continuous, and have varying boundaries. An individual is associated with a single set of parameters and it's corresponding figure of merit (FoM). It is also possible to define simulation parameters. The simulation parameters are not optimized themselves, but depend on one or more optimization parameters. The dependency is described by a mathematical expression, parsed and evaluated by an implementation of the Shunting-yard algorithm \cite{Shuntingyard}. 

In the main optimization loop of all algorithms, the optimization progress of each iteration is logged. When the optimization is completed, the best FoM and corresponding parameter values are put in a result file, along with the total number of evaluations, and information regarding the optimization mode. The result file also contains information about each iteration, including the number of evaluations required, best FoM found so far, and corresponding parameter values. Optionally, a file containing the best FoM of each iteration (as opposed to overall best FoM) and corresponding parameter values can be created. Additionally, the user can choose to trace any number of individuals and save all parameter values assumed by those individuals.

In all of the implemented optimization algorithms, the optimization parameters are initialized randomly within their individual boundaries. If an updated $\omega$ is outside of its boundaries at any point of the optimization process, it is shifted to the closest boundary by all algorithms. 

Discreet optimization parameters are treated as continuous throughout the optimization, and truncated before simulation, as suggested in Ref. \cite{DEmixedinteger}. After each simulation, the simulation result is used to calculate the FoM. 

\subsection{Particle Swarm Implementation}
PSO is a type of machine-learning algorithm which has its origins in the swarming of social animals, such as the schooling of fish or flocking of birds. 
A swarm consists of a set of candidate solutions called particles (refereed to as individuals here), which are characterized by their positions and velocities in time.
Each individual is aware of the best position it has seen and also the best position seen by any member of the swarm. When an individual $i$ moves in PSO, its new position is calculated by adding its velocity, 
$v_{i}$, to its position, $\omega_{i}$ and the velocity in a certain direction, $v_{i,j}$, is calculated as
\begin{center}
$v_{i,j}=c_{I}\cdot v_{i,j}+w_{L}\cdot$rand$(0,1)\cdot(\omega_{ibest,j}-\omega_{i,j})+w_{C}\cdot $rand$(0,1)\cdot(\omega_{best,j}-\omega_{i,j})$
\end{center}
where $\omega_{i,j}$ is the individual's position in direction $j$, and $\omega_{ibest,j}$ is the position in direction $j$ which has produced the best FoM so far for individual $i$. $\omega_{best,j}$ is the position in direction $j$ that has produced the overall best FoM for any individual. The inertial constant $c_{I}$, the local search weight $w_{L}$, and the collective search weight $w_{C}$, are all specified by the user and decide how greatly an individual's movement is influenced by it's own independent movement and that of the swarm. The random values of the equation are used in order to make the movement less predictable.
\subsection{Genetic Algorithm Implementation}
GAs are evolutionary algorithms which work on a population of individuals, which are selectively bred, to improve the quality of the population. When selecting individuals for reproduction in GA, two selection methods are available. Rank selection sorts the individuals from best to worst FoM and selects parents for reproduction with a probability relative to their position, making successful individuals more likely to reproduce. In tournament selection, two individuals are selected randomly, and the one with the best FoM is chosen for reproduction. 

The probability of reproduction is determined by the sexual rate, $r_{s}$, which is specified by the user. There is also a probability of the resulting individual being mutated
whether reproduction is done or a selected individual is copied over to the next generation. The mutation rate, $r_{m}$, is also specified by the user. Mutation is achieved by altering an individual's chromosomes, which are parameter values converted to Gray codes \cite{Graycodes}. During mutation, random bits of the chromosome Gray codes are inverted. The user can specify whether or not the GA should use elitism. When used, the best individual of each generation automatically continues on to the next generation. 

\subsection{Artificial Bee Colony Implementation}
ABC is a population-based algorithm which was inspired by the behavior of honeybees. In the algorithm, individuals are referred to as food sources.
In each iteration of ABC, a random parameter $\omega$ of each exploited food source will be mutated, by use of another randomly chosen food source. A mutated food source is compared to the original, and the food source resulting in the best FoM will be kept. 
The probability of a food source being selected for exploitation is proportional to the ratio of its FoM and the overall best FoM. Whenever a food source is selected, the mutation process is applied.
If a food source is not improved in a fixed amount of trials, it will be abandoned. This trial limit, $t_{l}$ is defined by the user. In case of abandonment, a new food source is initialized randomly. 
\subsection{Differential Evolution Implementation}
DE is a type of evolutionary algorithm which uses operations such as crossover, mutation, and selection to create a new generation of individuals from a population of previous ones. 
In each iteration of DE, the individuals of the population are mutated. The probability of an individual's optimization parameter being mutated is governed by the crossover constant, $c$. Mutated $\omega$ are calculated as
\begin{center}
$\omega_{i,j}=\omega_{r1,j}+m\cdot(\omega_{r2,j}-\omega_{r3,j})$
\end{center}
where $\omega_{r1,j-r3,j}$ belong to three randomly chosen individuals, such that  $\omega_{i}\neq\omega_{r1}\neq\omega_{r2}\neq\omega_{r3}$. The extent to which $\omega$ will be mutated is decided by the mutation constant, $m$. Both crossover and mutation are defined by the user.
The mutated individuals are compared to the original, and the individuals resulting in the best FoM are kept.

\section{Neutron optics and shielding applications}

The four optimization algorithms PSO, ABC, GA and DE have been applied to different neutron optics and shielding problems modeled in VITESS and Geant4. The performance of an algorithm was characterized by running 30 optimizations with 150 iterations and 50 agents for each problem and each algorithm. In the ABC optimizations, the maximum number of iterations was set to 75 instead of 150 because ABC uses twice as many evaluations per iteration. The mean best FoM as a function of evaluations was compared, as well as the mean and standard deviation of the finally reached FoM, the mean number of evaluations 
to find the best FoM ($N_{eval}$) and the maximum number of iterations without improvement before the final FoM was found ($N_{it}^{0}$). The latter is useful when deciding on a sensible non-improvement limit, i.e. an optimization can be stopped after a certain number of iterations without improvement in order to save unnecessary simulation time. Additionally, we also calculated the mean number of evaluations to get within 10$\%$ of the best FoM. This parameter is an indication
of how early in the optimization process a solution close to the best FoM is found. Table 1 lists the algorithm parameters and their values as used in the optimization runs.

\subsection{Neutron optics optimization results}

The optimization algorithms have been tested on two different kinds of neutron optics problems: a \textit{simplified example} with only two optimization parameters and a theoretically known solution, and a more \textit{realistic example} with 13 optimization parameters. 

\subsubsection{Simplified example}

This example is similar to the one used in the VITESS characterization of the built-in optimization routines\footnote{Two gradient methods, a metropolis algorithm and PSO are available since VITESS release 3~\cite{Vitess3} and 3.2, respectively.}: it contains a neutron source, an elliptically focusing guide and a pinhole in the focal plane of the ellipse. The optimization algorithms had to find the ideal position of the 1\,cm diameter pinhole in the horizontal and vertical direction, which was theoretically known to be (0/0). Gravity effects were switched off in this example. 
The FoM was the (negative of the) neutron intensity at the end of the simulation, i.e. the optimization maximized the neutron transport through the pinhole.

We also investigated a slight modification of the above example which was to introduce a plate with two pinholes, one at (0/0) and one at (1\,cm/1\,cm) and both with a diameter of 1\,cm between the guide exit and the focal plane. The resulting intensity distribution now contained a local maximum at (1\,cm/1\,cm) in addition to the global one at (0/0).  Furthermore, the generated source statistics in the Vitess simulation were reduced by a factor of 10. Interestingly, none of the algorithms got stuck in the local optimum in any of the 30 optimization runs, but all algorithms took more evaluations to find the vicinity of the global optimum.

Figure~\ref{f_NO_FoMvsEval} shows the development of the mean best FoM as a function of the mean number of evaluations for the two described scenarios. Table~\ref{t_NO_SimplifiedExample} summarizes the actual values and spreads of the best FoMs, as well as 
as the average number of evaluations until the best FoMs were found and also the average number of evaluations to get within 10$\%$ of these FoMs. All four algorithms delivered almost the same FoM. 
DE found the final best mean FoM, required the least number of total evaluations to find it and located the exact same solution for all 30 runs. ABC on the other hand found a similar FoM but needed more evaluations overall. 
PSO required the least number of evaluations to get within 10$\%$ of the best FoM, however this value was lower than both FoMs found by ABC and DE. The GA algorithm performed the worst and delivered the highest mean FoM with the largest spread
and also took the longest to come within 10$\%$ of the best FoM. 

If a local optimum was present in addition to the global one, the ABC and PSO algorithms improved the FoM fastest in the beginning of the optimization, but both took more evaluations to get to the final best FoM compared to the DE algorithm. In this case, ABC found a solution with the exact same FoM as the DE algorithm. While all algorithms improved the FoM much slower in the presence of a local optimum, GA seems to have suffered the most and has not completely converged at the end of the 150 iterations. 

Table~\ref{t_NO_SimplifiedExample} also lists the mean number of iterations without improvement before the absolute best FoM was found. This parameter can be useful for optimization problems as a user could define a non-improvement limit for the FoM. If during the optimization, a given number of iterations were carried out without improving the FoM, and this exceeded the user selected non-improvement limit, the optimization could be stopped in order to save unnecessary computational time. However, if the non-improvement limit was set to too small a value, the optimization would not have found the best FoM. For instance in this test example, the DE algorithm showed the smallest $N_{it}^{0}$, so it was not only the fastest algorithm to find the best FoM, but the optimization time could also be shortened more efficiently by setting a non-improvement limit.

Based on the above results, the DE and ABC algorithms appear to be the most suitable for the simple neutron optics optimization. ABC delivered a better result in the very beginning of the optimization process, but took a much longer time to reach the final result. 
PSO followed closely ABC in the beginning of the optimization but was unable to locate the same best FoM and also exhibited a higher spread of FoMs. The GA algorithm was the least suitable for this kind of optimization problem.

\subsubsection{Realistic example}

The more realistic example consisted of a doubly curved guide with straight or possibly linear sections in between, before and after the two curved sections. The curvature was in the horizontal direction with the second bent section in the opposite direction to the first, i.e. the whole guide built an S-shape. In the vertical dimension, the guide was a straight guide with constant width. While the guide entry and exit width were fixed, the width and length of the curved sections were variable. A schematic drawing of the model and its parameters are shown in figure~\ref{f_NO_RealisticExample}, with the fixed parameters marked in black, optimization parameters in red and simulation parameters, which were calculated from the optimization parameters, in green. The radius of curvature of each of the curved guide sections was calculated such that each section left direct line of sight. The formulas used for the calculations are given in Fig. 3. The coating of the left and right guide walls were set to be equal. With all guide coating values as well as entry and exit width of guide sections, the total number of simulation parameters calculated from the 13 optimization parameters was 14. Note that the length of the last guide section could be calculated to a negative value: in this case, the simulation failed and the FoM was set to 0. Therefore, this situation does not prevent the optimization from being successful under the condition that there are enough agents. The maximum m-value coating for the guides was set to 6.\\
\indent The FoM was set to be the (negative of the) signal over background at the guide exit, where the signal neutrons were defined as $\lambda_{signal}>$2\,\Angst\,and the background neutrons as $\lambda_{noise}<$2\,\Angst. However, the optimization results, and most steps in between, were in a phase space region where no background neutrons were transported to the end of the guide at all. In this case, the signal intensity was taken as the FoM in the optimization.  \\
\indent Figure~\ref{f_NO_FoMvsEvalB} shows the evolution of the FoM with the number of evaluations. It can be seen that the PSO algorithm performed best in the beginning of the optimization, but was surpassed by the DE algorithm which gave the best final FoM. 
It can be  seen from the data in Table 3 that both PSO and DE reached to within 10$\%$ of their best FoMs in roughly the same number of evaluations, however DE was performing better at this point.
Both the ABC and GA algorithms delivered significantly worse results. The spread in the final result was the same for all algorithms, as shown in table~\ref{t_NO_RealisticExample}.  \\
\indent In the overall best solution, the guide with fixed entry and exit width of 3 cm is expanded to 7 cm and 5.3 cm in the curved sections, choosing an overall ballistic shape at the expense of a higher curvature. 
It was found that two and one channel guides in the first and second curved sections with high supermirror coatings of around 5.2 were preferred to 
beam benders with a high number of channels and lower coating values. The optimization parameters, as indicated in Figure 3,
for the best solution were found to be: w$_{c1}$=6.94518 cm, L$_{lin1}$=84.0839 cm, m$_{lin1}$=5.98846, L$_{c1}$=8.5 m, Nch$_{c1}$=2, m$_{c1}$=5.26477, w$_{c2}$=5.28894 cm, L$_{lin2}$=10 cm, m$_{lin2}$=4.08117 cm, L$_{c2}$=7.0 m, Nch$_{c2}$=1, m$_{c2}$=5.21391, m$_{lin3}$=6, while the simulation parameters were: L$_{lin3}$=155.9 cm, R$_{c1}$=129.9 m, and R$_{c2}$=115.5 m. \\
\subsection{Shielding optimization results}
The Geant4 simulations were carried out using Geant4 version 10.0. The physics list used for the simulations was QGSP$\_$BERT$\_$HP \cite{Geant4physics} where ``QGS'' stands for the quark-gluon string model, ``P'' for precompound, ``BERT'' 
for the Bertini intra-nuclear cascade \cite{Bertini1963,Bertini1969}, and ``HP'' for the high-precision neutron tracking model for neutron interactions below 20 MeV. The optimization algorithms were tested on a problem based on the
design of a multi-layered beamstop or collimator block along the length of a neutron guide. In order to increase the computational efficiency of the simulations, geometrical splitting techniques were applied to the individual mass geometries of the multi-layered
structure.  \\
\indent The test problem included a multi-layered structure of up to 5 shielding blocks. Each individual block was allowed to vary between 0 and 5 cm, amounting to a total thickness of 25 cm. The blocks had a diameter of 1 meter. 
The material of any of the shielding blocks could be either concrete, paraffin, polyethylene, Al, Fe, Cu, Pb, or vacuum. The standard Geant4 materials were used in all cases. A 10 MeV neutron beam was incident on the first block and 
the number of neutrons exiting the multi-layered structure were recorded along with the average energy of those neutrons. \\
\indent The definition of the FoM was a critical parameter for this design scenario. The main goal of the multi-layered structure was to both minimize the number of neutrons emanating from the back surface and the energies
of those neutrons. Thus, the FoM was selected to be the number of emitted neutrons times the average energy of the neutrons. In practice, however, it may not be necessary to build the highest-performing solution or it may simply not be possible, due to budget
or floor-loading constraints, for example. To illustrate the effect of such a constraint, we also included a limit on the weight of the entire shielding structure. 
We arbitrarily set the weight limit and any solution above this value was given an infinitely large FoM. Solutions which were below this limit were allowed to keep the FoM defined as mentioned above.  \\
\indent The results of the optimization runs are highlighted in Figure 5 and Table 4. The mean FoM as a function of the mean number of evaluations is shown in Figure 5 and the final results are shown in Table 4. In the allotted simulation time, DE was able to find the lowest FoM on average and with the lowest spread while GA on average performed the worst with the highest FoM and largest spread. DE however took the longest number of evaluations 
to find the best FoM, required the third most number of evaluations to get within 10$\%$ of the value and also exhibited a high non-improvement limit. PSO took the least number of evaluations to come within 10$\%$ of its best FoM, but the value was
slightly worse than the FoM found by DE. \\
\indent The best overall solutions for the four different algorithms are shown in Table 5. All four algorithms found the best solutions to contain 3 blocks of metal shielding followed by two blocks of either paraffin or polyethylene.
In the case of PSO, GA, and DE, the ordering of the first three blocks was found to Fe, Cu, Fe and the thicknesses of the blocks were found to be roughly the same. In the case of ABC however, the optimization runs were unable to locate the Fe-Cu-Fe solution. \\
\indent The results presented here also highlight a major challenge when applying optimization algorithms in particular to shielding applications. The shielding example investigated included a simple geometry and beam description. For the DE algorithm, the average optimization took around 45 hours on a single CPU and the results in Figure 5 indicate that the FoM is still improving. The main time limiting factor is due to the Monte-Carlo simulations. This means that optimization of a realistic shielding example, including an entire beam length, 30 m for a short neutron scattering instrument and around 150 m for a long instrument, with all individual components and full energy spectrum will be extremely challenging. Advanced variance reduction and multiprocessing/multithreading techniques can be considered critical for this purpose. \\

\subsection{Conclusions and Summary}
From the results shown above, it can be seen that DE provided the best overall average FoMs with the lowest spreads for the neutron optics and shielding scenarios investigated in this work. 
However, it was also seen that the performance of the algorithms also depended on the specific problems. This likely was a result of the different FoM surfaces for each problem, which arise from a complex interplay of the free parameters, their limits, the definition of the FoM, statistical noise, and in these cases the underlying physics involved. Thus, it can be recommended to investigate a specific problem with DE and at least one other optimization algorithm.\\
\indent In summary, we developed software for exploring the benefits of metaheuristic optimization methods in neutron optics and shielding calculations. We tested several popular algorithms, including PSO, GA, ABC, and GA on three different design scenarios. Overall, we found that DE found on average the best solutions of the four algorithms and recommend the use of this algorithm in addition to one other algorithm in future applications of these techniques to neutron optics and shielding calculations. \\


\section*{References}
\bibliographystyle{elsarticle-num}
\bibliography{opti_paper}

\cleardoublepage
\begin{table}[h]
\caption{Algorithm parameters used in optimization as described in the text (For the ABC case, $D$ is the number of agents and $\omega$ is the number of parameters).}
\begin{center}
\begin{tabular}{|c|c|c|c|c|c|c|c|}
\hline
\multicolumn{3}{ |c| }{PSO} &
\multicolumn{2}{ |c| }{GA} &
\multicolumn{1}{ |c| }{ABC} &
\multicolumn{2}{ |c| }{DE}\\
\hline
$c_{I}$ & $\omega_{L}$ & $\omega_{C}$ & $r_{s}$ & $r_{m}$ & $t_{l}$ & $c$ & $m$\\ \hline
0.95 & 1.0 & 1.0 & 0.9 & 0.001 & 0.5$\times$D$\times\omega$ & 0.8 & 0.5\\ \hline
\end{tabular}
\end{center}
\label{default}
\end{table}%

\begin{table}[htb!]
\centering
\resizebox{\columnwidth}{!}{
\begin{tabular}{||c|c||c|c|c|c||}
\hhline{======}
                              & l.O.  & PSO         & GA           & ABC            & DE \\
\hhline{======}
-FoM                     & no &  740718$\pm$3  & 740693$\pm$5 & 740741$\pm$1 & 740745$\pm$0 \\
 $\left[\mathrm{100\,s}^{\mathrm{-1}} \right]$   & yes  & 740506$\pm$80  &   733108$\pm$4205 & 740931$\pm$0 & 740931$\pm$0 \\
\hline
$N_{eval}$                        & no   & 4847$\pm$282     & 4718$\pm$385      & 6167$\pm$216 & 4622$\pm$143 \\
                                 & yes  & 4600$\pm$315     & 5676$\pm$274      & 5508$\pm$114 & 2463$\pm$36 \\
\hline
$N_{eval}^{10\%}$                        & no   & 168$\pm$10     & 194$\pm$24      & 183$\pm$9 & 177$\pm$14 \\
                                 & yes  & 245$\pm$20     & 548$\pm$136      & 237$\pm$19 & 332$\pm$24 \\
\hline
$N_{it}^{0}$                     & no   & 44$\pm$4    & 38$\pm$5      & 17$\pm$2 & 21$\pm$2 \\
                               & yes  & 39$\pm$6    & 33$\pm$4      &  8$\pm$1 & 7$\pm$1 \\

\hhline{======}
\end{tabular}
}
\caption{Optimization results for the \textit{simplified example}: Mean best FoM, mean number of evaluations $N_{eval}$ until the best FoM was found, mean number of evaluations $N_{eval}^{10\%}$ to get to within 10$\%$ of the best FOM and number of iterations without improvement $N_{it}^{0}$ before the FoM was found, averaged over 30 optimization runs and as described in section 3.1.1, with and without local Optimum (l.O.).} 
\label{t_NO_SimplifiedExample}
\end{table}

\begin{table}[htb!]
\centering
\resizebox{\columnwidth}{!}{
\begin{tabular}{||c||c|c|c|c||}
\hhline{=====}
                                          & PSO & GA & ABC & DE \\
\hhline{=====}
-FoM [10$^{\mathrm{8}}$\,s$^{\mathrm{-1}}$]  & 8.9$\pm$0.3 & 7.7$\pm$0.2 & 8.0$\pm$0.2 & 10.6$\pm$0.2 \\
$N_{eval}$        & 6017$\pm$208 & 6885$\pm$41 & 7147$\pm$75 & 7153$\pm$73 \\
$N_{eval}^{10\%}$        & 3223$\pm$286 & 3280$\pm$212 & 4183$\pm$288 & 3233$\pm$123 \\
$N_{it}^{0}$ & 30$\pm$3 & 21$\pm$3 & 9$\pm$1 & 18$\pm$1 \\

\hhline{=====}
\end{tabular}
}
\caption{Optimization results for the \textit{realistic example}: Mean best FoM, mean number of evaluations $N_{eval}$ until the best FoM was found, mean number of evaluations $N_{eval}^{10\%}$ to get to within 10$\%$ of the best FOM and number of iterations without improvement $N_{it}^{0}$ before the FoM was found, averaged over 30 optimization runs.}
\label{t_NO_RealisticExample}
\end{table}

\begin{table}[t!]
\centering
\resizebox{\columnwidth}{!}{
\begin{tabular}{||c||c|c|c|c||}
\hhline{=====}
                                          & PSO & GA & ABC & DE \\
\hhline{=====}
FoM [neutrons*MeV]   & 435$\pm$3     & 473$\pm$10   & 460$\pm$4 & 410$\pm$3 \\
$N_{eval}$           & 4303$\pm$360    & 3968$\pm$198    & 5903$\pm$240 & 6407$\pm$182  \\
$N_{eval}^{10\%}$           & 1623$\pm$165    & 1962$\pm$109    & 3643$\pm$248 & 2957$\pm$215  \\
$N_{it}^{0}$           & 36$\pm$5     & 27$\pm$3     & 19$\pm$1   & 45$\pm$4 \\

\hhline{=====}
\end{tabular}
}
\caption{Optimization results for the \textit{Geant4 shielding example}: Mean best FoM, mean number of evaluations $N_{eval}$ until the best FoM was found, mean number of evaluations $N_{eval}^{10\%}$ to get to within 10$\%$ of the best FOM and number of iterations without improvement $N_{it}^{0}$ before the FoM was found, averaged over 30 optimization runs.}
\label{t_NO_RealisticExample}
\end{table}

\begin{table}[t!]
\centering
\resizebox{\columnwidth}{!}{
\begin{tabular}{||c||c|c|c|c||}
\hhline{=====}
                                          & PSO & GA & ABC & DE \\
\hhline{=====}
FoM [neutrons*MeV]       & 395         & 390         & 422         & 369        \\
Layer 1 (cm)               & 4.19 (Fe)   & 4.94 (Fe)   & 4.73 (Fe)   & 4.79 (Fe)  \\
Layer 2 (cm)               & 4.85 (Cu)   & 4.62 (Cu)   & 5.00 (Fe)   & 4.80 (Cu) \\
Layer 3 (cm)               & 4.94 (Fe)   & 4.78 (Fe)   & 5.00 (Fe)   & 4.71 (Fe)  \\
Layer 4 (cm)               & 5.00 (Para) & 4.84 (Poly) & 5.00 (Poly) & 4.83 (Para) \\
Layer 5 (cm)               & 4.86 (Para) & 4.81 (Para) & 5.00 (Para) & 4.98 (Para) \\

\hhline{=====}
\end{tabular}
}
\caption{The best overall solutions for the four algorithms applied to the Geant4 multi-layered shielding example. Para stands for paraffin and Poly for polyethylene. }
\label{t_NO_G4}
\end{table}

\cleardoublepage
\begin{figure}[!t]
\centering
\includegraphics[width=70mm]{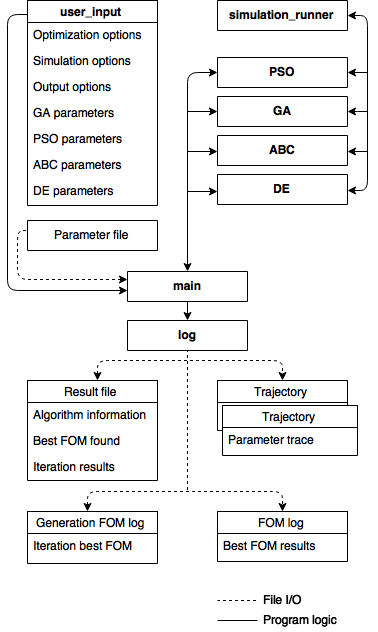}
\caption{Simplified visualization of software relations, focusing on input and output.}
\label{fig:flowchart}
\end{figure}

\begin{figure}
\includegraphics[width=0.95\linewidth]{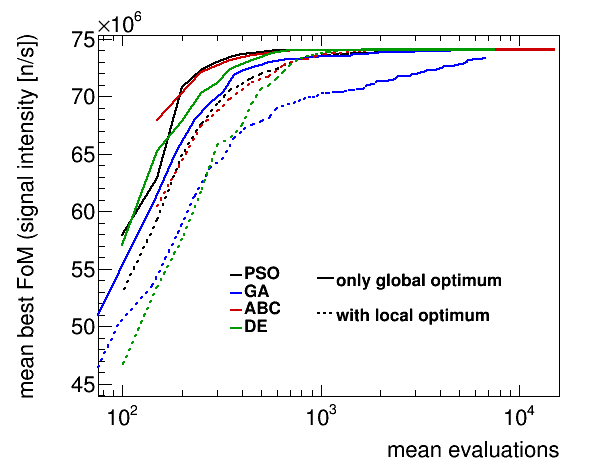}
\caption{Mean best FoM as function of the mean number of evaluations, averaged over 30 optimization runs for each algorithm. Simplified optimization problem with (dotted) and without (solid) local optimum.}
\label{f_NO_FoMvsEval}
\end{figure}

\begin{figure}[t!]
\includegraphics[width=\linewidth]{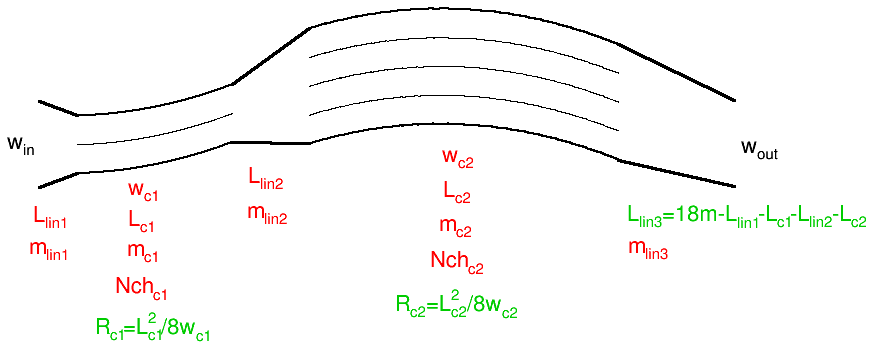}
\caption{Example of s-bender with linear guide sections: optimization parameters are shown in red, calculated simulation parameters in green and fixed parameters in black.
The parameter w is the guide width, L the length of the guide sections, m the supermirror coating, Nch the number of channels in the curved guide sections and R their curvature radius.}
\label{f_NO_RealisticExample}
\end{figure}

\begin{figure}[t!]
\includegraphics[width=0.95\linewidth]{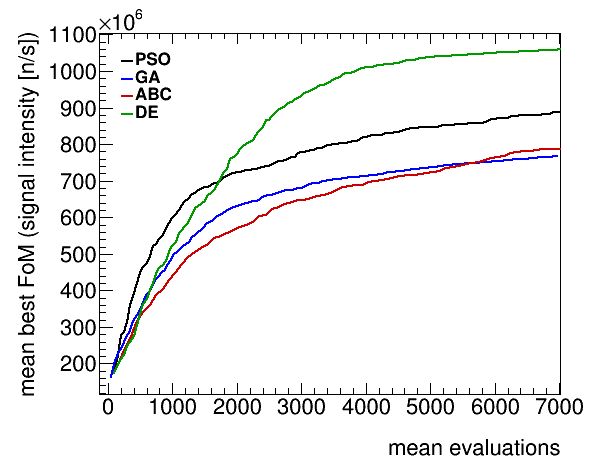}
\caption{Mean best FoM as function of the mean number of evaluations, averaged over 30 optimization runs for each algorithm, for the example with 13 optimization parameters.}
\label{f_NO_FoMvsEvalB}
\end{figure}

\begin{figure}[t!]
\includegraphics[width=\linewidth]{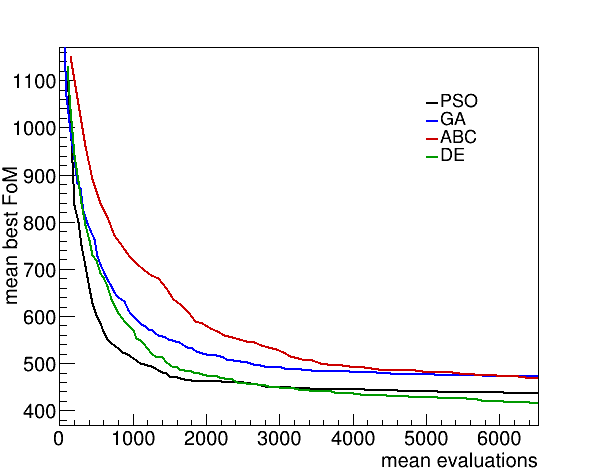}
\caption{Results of the Geant4 optimization for the multi-layered structure example.}
\label{f_Geant4_1}
\end{figure}

\end{document}